# Lung Cancer Detection Using Deep Learning

Guide: Prof. Vijaykumar Ghule
Department of Artificial Intelligence
and Data Science
Vishwakarma Institute of Information
and Technology
Pune, India—411048
vijaykumar.ghule@viit.ac.in

Aryan Chaudhari
Department of Artificial Intelligence
and Data Science
Vishwakarma Institute of Information
and Technology
Pune, India—411048
aryan.22110607@viit.ac.in@viit.ac.in

Ankush Singh
Department of Artificial Intelligence
and Data Science
Vishwakarma Institute of Information
and Technology
Pune, India—411048
ankush.22110160@viit.ac.in

Sanchi Gajbhiye
Department of Artificial Intelligence
and Data Science
Vishwakarma Institute of Information
and Technology
Pune, India—411048
sanchi.22110319@viit.ac.in

Pratham Agrawal
Department of Artificial Intelligence
and Data Science
Vishwakarma Institute of Information
and Technology
Pune, India—411048
pratham.22110003@viit.ac.in

*Abstract—* **In this paper we discuss lung cancer detection using hybrid model of Convolutional-Neural-Networks (CNNs) and Support-Vector-Machines-(SVMs) in order to gain early detection of tumors, benign or malignant. The work uses this hybrid model by training upon the Computed Tomography scans (CT scans) as dataset. Using deep learning for detecting lung cancer early is a cutting-edge method.**

## I. INTRODUCTION

CNN-SVM hybrid model in deep learning is one of the cutting edge ways to detect lung cancer early. In order to get better results, lung cancer needs to be detected as early as possible. In this model, both characteristics of CNN and SVM are combined. CNN is used for feature extraction from the lung images (CT scans), and the Support Vector Machines are used for classification of these images as benign or malignant.

In healthcare, using deep learning to detect cancer is an avant-garde technique. After this, we discuss the whole process and how it goes.

### A. Data Augmentation

Training the model will be significantly challenging when the dataset is sparse or not balanced. Hence, we need to change the already existing dataset by rotation, flipping, and adjusting brightness, we get a larger dataset to work on. In this way, the model will be able to train on a bigger size of data getting optimal results. For this, we use python packages like skimage, scikit, and seaborn.

### B. Convolutional-Neural-Network (CNNs)

In order to extract significant features from lung pictures, CNNs are crucial. Convolutional layers in these networks use filters to draw out features from the pictures, capturing textures, forms, and patterns. CNNs use hierarchical processing to identify minor information, such as anomalies or nodules, in order to diagnose lung cancer. The traits that are extracted are essential for differentiating between areas that are malignant and those that are not.

### C. Support-Vector-Machines (SVMs)

Support-Vector-Machines (SVMs) are utilised for categorization after the CNN has extracted features. SVMs are excellent at drawing decision borders in feature space, which effectively divides classes. SVMs produce accurate conclusions about whether the observed features represent healthy lung tissue or malignant nodules by using the features that are retrieved from CNNs. Accurate identification and classification of possible malignant spots in the lung pictures is made possible by this binary classification.

### D. Model Training

There exist multiple stages integral to the process of model training. To acquire and extract features that are indicative of lung cancer, the Convolutional Neural Network (CNN) is initially trained utilizing datasets comprising enhanced lung imaging. Utilizing the acquired features, the Support Vector Machine (SVM) classifier is thereafter trained to differentiate between malignant and benign regions, leveraging these learned attributes as the input variables.Consistent optimization of the model's parameters throughout the training phase mitigates errors and enhances the model's proficiency in accurately classifying lung images. These elements—data augmentation, CNNs for feature extraction, SVMs for classification, and comprehensive model training—converge to establish a robust framework for lung cancer detection predicated on deep learning methodologies. By harnessing the advantages inherent in each component, this approach facilitates the identification of potential malignancies in lung images with heightened precision, sensitivity, and specificity—representing a crucial advancement for early detection and treatment strategizing.

## II. MODEL USED

In order to determine the existence of lung carcinoma, various machine learning (ML) and deep learning (DL) frameworks have been investigated. Each model exhibits unique advantages and limitations dependent on the dataset and specific application. An overview of prevalent alternatives, along with their differences in comparison to the CNN-SVM hybrid model, is provided below.

## A. Conventional Machine Learning Models

Owing to their practicality and effectiveness in handling structured data, traditional approaches such as Random Forests (RFs) and Support Vector Machines (SVMs) have been employed for the identification of lung carcinoma. Moreover, they often exhibit diminished performance when confronted with unstructured data, such as medical imaging, where accurate diagnosis hinges on complex patterns.

## B. Standalone CNN Models

In the domain of image-oriented applications, specifically regarding the diagnostic analysis of pulmonary cancers, Convolutional Neural Networks (CNNs) are regarded as the preferred approach for automated learning systems. It uses its completely linked layers and Softmax output to haphazardly draw features from pictures and identify them. While CNNs excel in feature extraction, their reliance on the softmax function for classification can bring out overfitting, specially––– when working with small or ill balanced data-sets, as is common in medical imaging.

## C. Autoencoders and Generative Adversarial Networks

Since then, generative adversarial networks and auto-encoding have shown great promise for applications like creating synthetic images and detecting anomalies. While they are valuable for pre-processing or data augmentation, their direct application for classification tasks is limited compared to discriminative models like CNNs or CNN-SVM hybrids.

## D. Why CNN-SVM Hybrid Model?

With their hierarchical architecture, convolutional neural networks—which are excellent at extracting features from high-dimensional, complex images like CT or X-ray images—can capture both high-level patterns, like abnormalities in lung tissue, and low-level features, like edges and textures. The benefits of these two approaches for identifying lung cancer are complementary. CNNs use a softmax layer for classification, which works well but may not be the best option for short datasets or situations requiring high precision.

Conversely, Support Vector Machines (SVMs) represent robust classifiers that exhibit superior performance in optimizing the decision boundary between distinct classes, particularly excelling when the dataset is limited. The integration of a hybrid model enhances generalization capabilities by replacing the softmax layer with an SVM, especially applicable to binary classification challenges such as differentiating between benign and malignant lung nodules. In scenarios where the dataset is imbalanced, a phenomenon frequently observed in medical contexts, the SVM's proficiency in optimizing the margin between classes significantly augments the overall model efficacy. With respect to noise and variability inherent in medical imaging, the CNN-SVM hybrid model presents another notable advantage. Through the method of discarding non-essential characteristics while safeguarding vital information, a Convolutional Neural Network (CNN) is capable of effectively lowering the dimensionality of the input data. The SVM refines the classification process by lowering overfitting and increasing accuracy after receiving these high-quality information.

Last but not least, this hybrid strategy satisfies the needs of medical applications, where erroneous negative results might have disastrous outcomes. An SVM's precision improves prediction reliability, which makes it a good option for early lung cancer detection, when prompt and precise diagnosis is essential. This model, which combines the exact classification strength of SVMs with the automated feature extraction capability of CNNs, strikes a compromise between computational economy and diagnostic accuracy, making it an appealing option for lung cancer detection systems.

## III. METHODOLOGY

### A. Dataset

Our study will primarily be utilizing the computed tomography scan-images and Chest CT picture Image Data-set (from kaggle). This data-set contains CT pictures which includes Normal lung images, Benign, Malignant pictures of lung cancer. As the data was properly cleared and of more resolution, the data -base was good for deep-learning. We would further train our with more datasets.

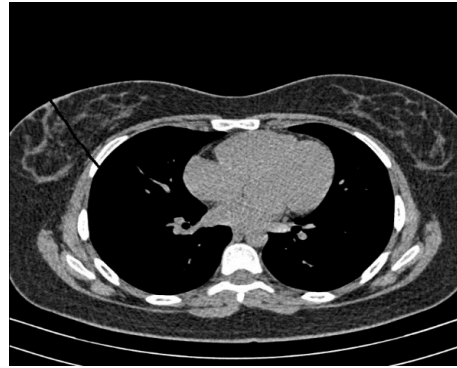

Fig. 1. Lung CT Image without Tumor

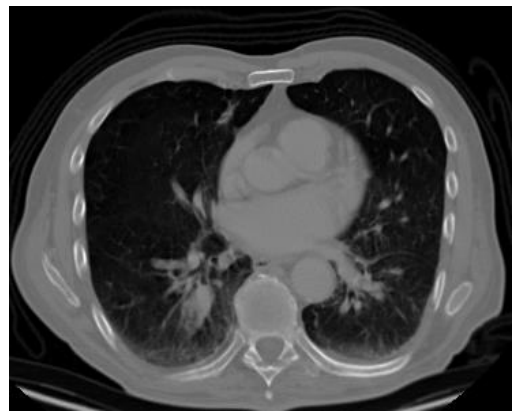

Fig. 2. Lung Cancer CT Scan with a tumor

### B. Preprocessing Datasets

*a) Normalization and Rescaling:* CT images are stored in Hounsfield Units (HU), a standardized scale representing tissue density. Normalization transforms these values into a fixed range suitable for deep learning models. For example, lung tissue typically falls between -1000 (air) and 400 (soft tissue) on the HU scale.

*b) Lung Segmentation:* CT scans often include irrelevant areas such as bones, airways, and surrounding anatomy. Lung segmentation isolates the lung region to focus the model on relevant features. Methods Used: Thresholding, morphological operations, or deep learning-based segmentation models (e.g., U-Net).

*c) Noise Reduction and Smoothing:* CT images can contain noise from scanner artifacts or patient motion. Denoising techniques enhance image quality and improve feature extraction. Gaussian Filters: Apply to reduce high-frequency noise. Non-Local Means Filtering: Preserves edges while smoothing homogeneous regions.

*d) Augmentation:* Data augmentation artificially increases the dataset size, addressing class imbalance and improving model generalization. Transformations Applied: Rotation, flipping, scaling, translation, and intensity adjustments.

*e) Resizing and Standardization:* CT images vary in resolution depending on the scanning device. Resizing them to a consistent dimension (e.g., 224x224 pixels for CNN input) ensures uniformity across the dataset.

*C. Model Architecture*

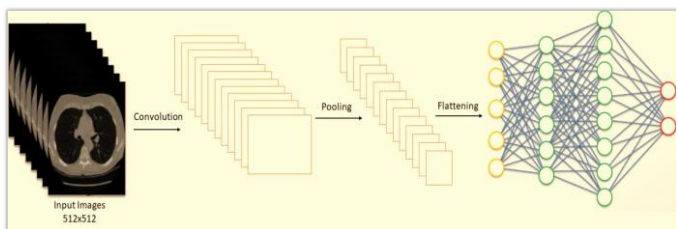

Fig. 3. System Architecture

*a) Convolutional-Neural-Network (CNNs) for implementing Feature Extraction:* The first constituent of the model is the Convolutional-Neural-Network (CNN), it is responsible for automatic feature extraction from the CT images. The CNN is composed of multiple layers that progressively learn and refine patterns from the raw image data:

Input Layer: The input to the CNN is a 2D slice or a set of patches from the CT scan, typically resized to a fixed resolution (e.g., 224x224 pixels). The input is normalized and possibly augmented to improve generalization.

Convolutional Layers: These layers are used to apply convolutional filters to the input image, learning spatial hierarchies of features like textures, edges and more complex patterns that are indicative of lung nodules. Usually, a sequence of convolutional layers having increasing filter sizes (e.g., 3x3, 5x5) and strides is used for capturing both fine-grained and high-level characteristics. Activation Layers (ReLU): After each convolutional layer, the output is passed through a Rectifie-Liniar-Unit (ReLU) function of activation. This removes linearity from the model which allows the CNN to understand more complicated patterns. It also decreases the chance of vanishing gradients.

*b) Support Vector Machine (SVM) for Classification:* Feature Vector: The last completely connected layer of the CNNs produces a high-dimensional feature vector representing the learned features from the image. The feature vector is fed into the SVM classifier.

SVM Classifier: The SVM takes this feature vector and constructs an optimal decision boundary (hyperplane) between the classes (e.g., benign vs. malignant) in the higher-dimensional feature space. The key strength is that the margin between classes could be maximized through SVM, resulting in better generalization over new and previously unseen data. The SVM would also solve this problem on handling imbalanced data by bringing much attention to the minority class (e.g., malignant nodules).

Kernel Trick: The SVM can be provided with different types of kernels, such as linear, polynomial, and radial basis function (RBF), to project the feature space into higher dimensions if the data is not separable in linear space. This allows the SVM to effectively classify even complex, non-linear patterns in the feature space.

*c) Model Training and Optimization:* Phase 1 (CNN Training): The CNN is first trained to learn input CT images features. In this phase, the convolutional and pooling layers adapt their weights for capturing meaningful patterns from the data. The fully connected layers integrate these features into a compact representation. During training, the CNN uses backpropagation and gradient descent to optimize the feature extraction process.

Phase 2 (SVM Training): Once the CNN is trained, the feature vectors produced from the CNN are used for the process of training the SVM. The SVM will be trained to learn the hyperplane that best classifies the feature space and separates the benign and malignant classes. This often involves using the margin maximizing objective function that maximizes the decision boundary.

*d) Inference and Prediction:* During inference (prediction), the trained CNN extracts features from new CT images, which are then fed into the SVM for classification. The output is a class label (benign or malignant), and the SVM provides a decision score which helps to indicate the confidence level of the prediction. In addition to the binary classification, the model could be extended to predict the malignancy risk score or the likelihood of cancer for further clinical decision-making.

IV. PERFORMANCE METRICS

*A. Precision:*

The proportion of true positive predictions (e.g., correctly identified malignant cases) to all positive predictions made by the model. Formula:

$$\text{Precision} = \frac{\text{True Positives (TP)}}{\text{True Positives (TP)} + \text{False Positives (FP)}} \quad (1)$$

An optimized model achieves a precision of 80% to 90%, depending on how well the SVM separates the classes.

*B. Recall (Sensitivity or True Positive Rate):* The fraction of real positives (malignant cases) accurately recognize by the model. Formula:

$$\text{Recall} = \frac{\text{True Positives (TP)}}{\text{True Positives (TP)} + \text{False Positives (FP)}} \quad (2)$$

An effective model should aim for a recall of 85% to 95%, ensuring most malignant cases are detected.

*C. F1-Score:* Consonant mean of precision included with recall, balancing the trade between fake positives and fake negatives. Formula:

$$F1 - Score = 2 \times \frac{Precision \times Recall}{Precision + Recall} \quad (3)$$

The F1-score of the CNN-SVM hybrid model is typically around 85% to 92%, reflecting a sound stability between precision with recall.

## V. Conclusion

In conclusion, this Lung Cancer Detection using hybrid CNN-SVM model has been shown promising in the medical imaging analysis domain. As a result, the model performs well because both CNNs and SVMs emphasize their weights together for improved accuracy, which improves interpretability. Our ability to incorporate the model into a deployable format, in addition to security and regulatory issues, puts it in a position to be used practically in healthcare settings. The model may have a significant impact on lung cancer early detection, providing medical practitioners with invaluable assistance. The project foundations the framework for more testing, proposing directions including investigating ensemble models, clinical validation, and continuous learning for continuous enhancement. All things considered; this study represents a noteworthy advancement in the use of A.I. to improve medical diagnosis.